\documentclass[pra,showpacs,showkeys,amsmath,amssymb,superscriptaddress,preprintnumbers,floatfix, preprint]{revtex4}
\usepackage{bm,amsfonts, mathtools}
\usepackage{graphicx,color, wasysym}
\usepackage{textcomp}

\begin{document}

\makeatletter

\title{On Mittag-Leffler function and associated polynomials}

\author{D. Babusci}
\email{danilo.babusci@lnf.infn.it}
\affiliation{INFN - Laboratori Nazionali di Frascati, via E. Fermi, 40, IT 00044 Frascati (Roma), Italy}

\author{G. Dattoli}
\email{dattoli@frascati.enea.it}
\affiliation{ENEA - Centro Ricerche Frascati, via E. Fermi, 45, IT 00044 Frascati (Roma), Italy}

\author{K.~G\'{o}rska}
\email{Katarzyna.Gorska@ifj.edu.pl}
\affiliation{H. Niewodnicza\'{n}ski Institute of Nuclear Physics, Polish Academy of Sciences, ul.Eljasza-Radzikowskiego 152, 
PL 31342 Krak\'{o}w, Poland}
\affiliation{ENEA - Centro Ricerche Frascati, v. le E. Fermi, 45, IT 00044 Frascati (Roma), Italy}

\begin{abstract}
The Mittag-Leffler function plays a role of central importance in the theory of fractional derivatives. In this brief note we discuss the properties 
of this function and its connection with the Wright-Bessel functions and with a new family of associated heat polynomials.
\end{abstract}

%--------------------------------------
\maketitle

%-----------------------------------------------------------

The Mittag-Leffler function (MLF) \cite{AAKilbas06, HJHaubold09} plays a central role in the theory of fractional derivatives (FD) \cite{AAKilbas06, IPodlubny99}. 
It has been thoroughly investigated \cite{IPodlubny99} but the increasing interest for FD in applications \cite{RMetzler99} demands for further studies, 
eventually leading to further properties or to more efficient methods of computation and analysis.

The importance of FD differential equations to model anomalous diffusion \cite{RMetzler99, EBarkai01} has been recognized in the past two decades when 
it has been shown that the evolution of the distributions associated with this process can be modeled by equations of the type \cite{TFNonnenmacher95, BJWest10}
\begin{equation}
\label{FFP}
\partial_t^\alpha\,F_\alpha (x, t) = \hat{L}_{FP}\,F_\alpha (x, t) + \frac{t^{- \alpha}}{\Gamma(1 - \alpha)}\,f(x), \qquad F_\alpha (x, 0) = f(x), 
\end{equation}
with $\alpha \in (0, 1)$, and $\hat{L}_{FP}$ being the Fokker-Plank (FP) operator associated with the specific problem under study. The use of the evolution 
operator method has been proved to be an efficient tool for searching the solution of this kind of problems \cite{GDattoli05}. Therefore, limiting ourselves to the case 
$\hat{L}_{FP} = k_\alpha\,\partial_x^2$, we can cast the solution of Eq. \eqref{FFP} in the form 
\begin{equation}
\label{FPpro}
F_\alpha (x, t) = E_\alpha (k_\alpha\,t^\alpha\,\partial_x^2)\,f(x), \qquad \qquad E_\alpha (y) = \sum_{n = 0}^\infty \frac{y^n}{\Gamma (1 + \alpha\,n)},
\end{equation}
where $E_\alpha (y)$ is the MLF. The operator $E_\alpha (k_\alpha\,t^\alpha\,\partial_x^2)$ is the evolution operator for the FP problem in Eq. \eqref{FFP} and the 
associated rules can be found in Refs. \cite{GDattoli05, GDattoli03}.

Within the framework of FD, the MLF plays the same role of the exponential function in the case of ordinary calculus. In the case $f (x) = x^n$, from Eq. \eqref{FPpro} we 
find \cite{GDattoli03} 
\begin{equation}
\label{Falf}
F_\alpha (x, t) = \,_{\alpha}H_n (x, k_\alpha\,t^\alpha) = n!\, \sum_{m = 0}^{[n/2]} \frac{x^{n - 2 m}\,(k_\alpha\,t^\alpha)^m}{(n - 2\,m)!\,\Gamma(1 + \alpha\,m)}, 
\end{equation}
where $_{\alpha}H_n (x, y)$ is the fractional heat polynomials of order $\alpha$. Their names derive from the so-called heat polynomials (sometimes Hermite heat polynomials), 
which are the natural solutions of the ordinary heat equation. The generating function of this family of polynomials is 
\begin{equation}
\sum_{n = 0}^{\infty} \frac{\xi^n}{n!}\,_{\alpha}H_n (x, y) = E_\alpha (\xi^2\,y)\,e^{\xi\,x},
\end{equation}
and, therefore, they are members of the Appell family \cite{PAppell26} satisfying the recurrences
\begin{align}
\partial_x\,_{\alpha}H_n (x, y) &= n\, \,_{\alpha}H_{n - 1} (x, y), \nonumber \\
\partial_y^\alpha\,_{\alpha}H_n (x, y) &= n\,(n - 1)\,_{\alpha}H_{n - 2} (x, y) + \frac{y^{-\alpha}}{\Gamma(1 - \alpha)}\,x^n.
\end{align}
To explore more in depth the nature of these polynomials, we consider a function $F(x)$ that can be expanded in terms of them (we assume that $y$ plays the role of a parameter)
\begin{equation}
F(x) = \sum_{n = 0}^\infty a_n\,\,_{\alpha}H_n (x, y), 
\end{equation}
with the coefficients $a_n$ to be determined. In the case of Eq. \eqref{Falf}, it can be shown that \cite{KAPenson10, KGorska11-1}
\begin{equation}
\,_{\alpha}H_n (x, t^\alpha) = \int_0^\infty \mathrm{d}s\,n_\alpha (s, t)\,\,_{1}H_n (x, s) \qquad\qquad 
n_\alpha (s, t) = \frac1{\alpha}\,\frac{t}{s^{\,1 + 1/\alpha}}\,g_\alpha\left(\frac{t}{s^{1/\alpha}}\right),
\end{equation}
with $g_\alpha (x)$ being one-sided L\'{e}vy stable distributions (see Eqs.(2), (3), and (4) in Ref. \cite{KAPenson10}). Therefore, we can write
\begin{equation}
\label{Fxexp}
F (x) = \int_0^\infty \mathrm{d}s\, n_{\alpha}(s, y^{1/\alpha})\,G(x, s)
\end{equation}
where
\begin{equation}
G(x, s) \,=\, \sum_{n = 0}^{\infty} a_{n}\, \,_{1}H_n (x, s).
\end{equation}
We can accordingly conclude that a function can be expanded in series of fractional heat polynomials whenever a corresponding expansion in series of ordinary heat polynomials exists.
The polynomials $_{\alpha}H_n (x, y)$ cannot be considered orthogonal, however the identity \eqref{Fxexp} establishes a close links with the ordinary case and with the relevant orthogonality 
properties. Further comments on these aspects will be presented elsewhere.

The function $E_\alpha (-x^2)$ can be considered as the generalization of Gaussian and its properties can be studied using an operational technique recently suggested in 
Refs. \cite{DBabusci11, KGorska11}. According to this technique, by defining an operator $\hat{c}$ such that
\begin{equation}
\hat{c}^{\,\alpha}\,\varphi(0) = \varphi(\alpha + 1), \qquad\qquad (\alpha \in \mathbb{R})
\end{equation}
for $\varphi (\mu) = 1/\Gamma (\mu)$ we can write
\begin{equation}
\label{Eal}
E_\alpha (- x^2) = \sum_{k = 0}^\infty (-1)^k\,x^{2\,k}\,\hat{c}^{\,\alpha\,k}\,\varphi (0) = \frac1{1 + \hat{c}^{\,\alpha}\,x^2}\,\varphi(0).
\end{equation} 
Albeit apparently trivial, if correctly used, last equation can usefully be exploited to draw many consequences. Let us, for example, consider the integral
\begin{equation}
I^{(2)}_{\alpha} = \int_{- \infty}^\infty \mathrm{d}x\,E_\alpha (- x^2).
\end{equation}
As a consequence of Eq. \eqref{Eal}, taking into account the identity
\begin{equation}
\frac1{A^\nu} = \frac1{\Gamma (\nu)}\,\int_0^\infty \mathrm{d}s\,e^{- s\, A}\,s^{\nu - 1}, 
\end{equation}
and treating $\hat{c}^{\alpha}$ as a constant, it's easy to show that
\begin{equation}
I_\alpha^{(2)} = \frac{\pi}{\Gamma \left(1 - \displaystyle \frac{\alpha}2\right)}.
\end{equation}
The same procedure can be used to evaluate successive derivatives of the function $E_{\alpha}(-x^2)$. From Eq. \eqref{Eal}, one has
\begin{equation}
\partial_x^n\,E_\alpha (- x^2) = \sum_{k = 0}^\infty \frac{(-1)^k\,(2\,k)!}{(2\,k - n)!\,\Gamma(1 + \alpha\,k)}\,x^{2\,k - n}.
\end{equation}

According to our formalism, the function defined by 
\begin{equation}
\label{Wrig}
W_{\alpha, \beta} (x) = \hat{c}^{\,\alpha - 1}\, e^{\hat{c}^{\,\beta}\,x}\,\varphi(0),
\end{equation}
has the following series expansion
\begin{equation}
W_{\alpha, \beta} (x) = \sum_{k = 0}^\infty \frac{x^k}{k!\,\Gamma(\alpha + \beta\,k)},
\end{equation}
i.e., it is the Wright function \cite{GEAndrews01,RGorenflo99}. By using the representation \eqref{Wrig} for this function, it is straightforward to derive the following 
identities
\begin{equation}
\partial_x^m\,W_{\alpha, \beta} (x) = W_{\alpha + m \beta, \beta} (x), 
\end{equation}
and
\begin{equation}
\int_{-\infty}^\infty \mathrm{d}x\,W_{\alpha, \beta} (-x^2) = \frac{\sqrt{\pi}}{\Gamma \left(\alpha - \displaystyle \frac{\beta}2\right)}.
\end{equation}

Let us now consider the family of polynomials of Laguerre. It is possible to define a fractional version of the two-variable type as follows
\begin{equation}
\,_{\alpha}L_n (x, y) = n! \,\sum_{k = 0}^n \frac{(-1)^k\,x^{\,\alpha\,k}\,y^{\,n - k}}{k!\,(n - k)!\,\Gamma (1 + \alpha\,k)}.
\end{equation}
These polynomials satisfy the differential equation
\begin{equation}
\label{dLag}
\partial_{y}\,F_{\alpha} (x, y) = \,_{\alpha}\hat{D}_{L, x}\,F_\alpha (x, y), \qquad\qquad F_\alpha (x, 0) = \frac{(-1)^n\,x^{\,n\,\alpha}}{\Gamma (1 + n\,\alpha)}.
\end{equation}
where the fractional Laguerre derivative $\,_{\alpha}\hat{D}_{L, x} = -(1/\alpha)\,\partial_x^\alpha\,x\,\partial_x$ has been introduced (for the notion of the ordinary 
Laguerre derivative see Refs. \cite{GDattoli04, GDattoli05-1}). This implies that the fractional Laguerre polynomials can be defined by means of the identity
\begin{equation}
\,_{\alpha}L_n (x, y) = e^{\,y\,_{\alpha}\hat{D}_{L, x}}\,\frac{(-1)^n\,x^{n\,\alpha}}{\Gamma (1 + n\,\alpha)},
\end{equation}
from which we find
\begin{equation}
\sum_{n = 0}^\infty t^n\,_{\alpha}L_n (x, y) = e^{\,y\,_{\alpha}\hat{D}_{L, x}}\,E_\alpha (-x^{\,\alpha}\,t).
\end{equation}
This generating function can also be obtained by replacing the exponential function with MLF in the generating function of the ordinary two-variable polynomials 
\cite{GEAndrews01}, and, therefore, we can write
\begin{equation}
\sum_{n = 0}^\infty t^n\,_{\alpha}L_n (x, y) = \frac1{1 - y\,t}\,E_\alpha \left(- \frac{x^{\,\alpha}\,t}{1 - y\,t}\right)
\end{equation}
from which we conclude that the function on the l.h.s. of this equation is the solution of Eq. \eqref{dLag} with initial condition $E_{\alpha}(- x^{\,\alpha}\,t)$. Furthermore, 
it is also easily proved that
\begin{equation}
\sum_{n = 0}^\infty \frac{t^n}{n!}\,\,_{\alpha}L_n (x, y) = e^{\,y\,t}\,W_{1, \alpha} (-x^\alpha\,t).
\end{equation}

%----------------------------------------------

\end{document}